\documentclass[12pt,preprint]{aastex}

\shorttitle{Twisted coronal loops in uniform gravity}
\shortauthors{Petrie}

\begin{document}


\title{On the expansion of quasi-static, twisted coronal loops in uniform gravity}


\author{G.J.D. Petrie}
\affil{National Solar Observatory, 950 N. Cherry Avenue, Tucson, AZ 85719}



\begin{abstract}

Coronal loop emission profiles are often of remarkably constant width along their entire lengths, contradicting expectations based on model coronal magnetic field strengths decreasing with height.  Meanwhile Paul Bellan has produced a theoretical model in which an initially empty, twisted force-free loop, on being filled with plasma via upflow at each foot point, in the absence of significant gravitational effects, forms a narrow, filamentary loop of constant cross-section.  In this paper, we focus on equilibrium states that include stratification by uniform gravity while retaining the effects of magnetic field twist.  Comparing these with related force-free equilibria, it is found that injection of low-$\beta $ plasma under coronal conditions is not likely to change the shape of a loop significantly.  These linear equilibria apply to the interiors and boundaries of loops only, with external influences modeled by boundary total pressures.  The effects of total pressure balance with surroundings and of gravitational stratification are to inhibit the pinching of a loop to a constant cross-section.  Only if the plasma $\beta$ were high enough for the plasma to reconfigure the external field and the hydrostatic scale height much greater than the loop size could the final state have nearly constant cross section.  We do not expect this to occur in the corona.

\end{abstract}



\section{Introduction}

The most emissive steady solar coronal fields at MK temperatures are the closed, loop-shaped structures that trap coronal plasma against thermal expansion (Rosner et al.~1978, Aschwanden~2004).  Understanding the basic physical properties of these structures is a crucial step towards understanding the corona.  Closed plasma loops are believed to trace lines of magnetic force that penetrate the photosphere from below and expand to fill the coronal volume above an active region (Bray et al.~1991, Aschwanden~2004).  They are curved filamentary structures connecting regions of opposite photospheric magnetic polarity.  Plasma loops are often observed as isolated brightened flux trajectories embedded in a less emissive field configuration.  A broad consensus has developed that coronal loops are nearly isothermal along their arclengths, while temperatures and densities can vary on small scales between neighboring loop structures (Rosner et al.~1978, Lenz et al.~1999, Aschwanden et al.~1999, 2000, Aschwanden and Nitta~2000, Schmelz et al.~2001, Martens et al.~2002, Schmelz~2002, Winebarger et al.~2002).  This can be explained by the thermal conductivity being much more efficient along than across the magnetic field in the nearly fully-ionized coronal gas (Spitzer~1962) and by the very high electrical conductivity of coronal plasma which "freezes" the plasma into the field lines of force (e.g. Aschwanden~2004).

However, much remains to be learnt about coronal loops.  Two physical parameters that continue to cause controversy are the scale height of hydrostatic stratification and the width of the observed loop-shaped emission patterns.  Loop emission in active regions is frequently clearly visible at heights several times greater than the hydrostatic scale height associated with the temperature of emission (Aschwanden et al.~2001).  Furthermore, the widths of these emission signatures remain remarkably constant along their entire lengths (Klimchuk~2000, Watko \& Klimchuk~2000).  These emission patterns are usually identified with magnetic flux tubes because of the efficient field-aligned thermal conductivity of the corona, and they therefore defy theoretical expectations that flux tubes must expand in a stratified magnetic field (L\'opez Fuentes et al.~2006).  In a divergence-free flux distribution (${\bf\nabla}\cdot{\bf B} =0$ where ${\bf B}$ is the magnetic field vector), flux tube cross-sectional area and field strength must be inversely proportional.  Since the macroscopic coronal magnetic field may be approximated by a superposition of magnetic multipoles (Altschuler \& Newkirk~1969) and there is obviously more space high in the solar atmosphere than low down, we expect the field strength to decrease with increasing altitude.  Pioneering direct measurements of coronal magnetic field strength by Lin et al.~(2004) also show a marked decrease of field strength with increasing altitude.  Yet loop emission profiles generally have near-constant width.   While concerns have been raised about the effects of background subtraction and spatial resolution on the measurement of the widths of these narrow structures, in practice width variations and background are not correlated and loops with constant cross-section can be distinguished with confidence from loops with expansion factors characteristic of force-free models, even if these widths are close to the resolution limit (L\'opez Fuentes et al.~2008).

The variation of cross-sectional area along a loop is particularly important to understanding plasma dynamics along loops.  Klimchuk et al.~(2000) showed that the tension force associated with
twist tends to make more twisted flux tubes more circular
in shape, but that the near-constant cross-sections of
loops cannot be explained by this twist.  Petrie et al.~(2003), Patsourakos et al.~(2004) and Gontikakis et al.~(2005) recently conducted different studies of how steady plasma flows along loops may influence the observed fall-off of plasma density, and therefore emission intensity, with increasing height.  Patsourakos et al.~(2004) assumed that loop cross-sectional areas are constant or nearly constant along loops, invoking the observations of Yohkoh/SXT and TRACE loop widths by Klimchuk~(2000) and Watko \& Klimchuk~(2000).  Traditionally HD models have assumed a constant cross-section (Orlando et al.~1995a, b; Reale et al.~2000a, b,  Peres~2000, Winebarger et al.~2003).  Cargill \& Priest~(1980, 1982) investigated the effects of flux tube cross-sections varying linearly with arclength.  Motivated by interplay between characteristic scale heights of plasmas and magnetic fields, Petrie~(2006b) investigated steady flows of stratified plasmas in flux tubes whose cross-sectional widths decrease exponentially with increasing altitude.  The outcome is that flux tube expansion can enable the extension of plasma scale heights by steady flows, and such flows can extend scale heights to observed scales, but that the conditions under which this can occur are rather special and are unlikely to be met generally in loop structures.

In this paper we address the issue of how loop-shaped emission patterns of constant cross-section are created.  In particular, we investigate whether or not a mechanism proposed by Bellan~(2003) for the creation of thin, filamentary flux tubes of constant cross-section, formulated in the absence of gravity, can work in the gravitationally stratified medium of the corona.  Bellan presented a sequence of physical processes which result in a thin, hot flux tube of exactly constant cross-sectional width. The model is consistent with the thin, filamentary structures of solar loop emission patterns as well as with various experiments performed by Bellan's plasma physics group (e.g. You et al.~2005). 

In axisymmetric geometry with cylindrical coordinates $(r,\theta ,z)$, where the symmetry axis $r=0$ coincides with the central axis of the loop flux tube, the process begins with a current-free loop, i.e., $B_{\theta}=0$, with field stronger at the foot points $z=\pm L$ than at the apex $z=0$.  This means that the tube is wider at the apex than at the ends initially.  If a current $I$ flows along this tube then a toroidal field component $B_{\theta}\ne 0$ is introduced and the field becomes twisted.  The resulting non-conservative Lorentz (${\bf j}\times{\bf B}$) forces then accelerate plasma from the narrow ends of the tube towards the apex.    In Bellan's~(2003) picture, these flows thereby advect frozen-in toroidal magnetic flux from each foot point towards the apex so that there is excessive magnetic tension there.  This tension pinches the loop until a new equilibrium is found.  In the absence of gravity the only possible equilibrium is one-dimensional and independent of $z$: it has constant plasma and magnetic flux distributions along the entire length of the loop and so we arrive at a thin loop of constant cross-section.

An important difference between the solar case and the case studied by Bellan~(2003) is the presence of a significant gravitational force in the solar corona.  While the coronal gas is not dense, the gravitational scale height of plasma at typical coronal temperatures, $\approx 1$~MK, is  about 60~Mm, where we assume a fully ionized hydrogen plasma.  Aschwanden et al.~(2001,2004) quote a value of about 47~Mm, assuming a 10:1 ratio of hydrogen to helium.  Therefore, loops of active region scales, of order 100,000~km, should be modeled with gravitational effects taken into consideration.

The dynamics and the final equilibrium problem described above are affected by the introduction of gravity.  The acceleration of plasma from the thin parts of the loop towards the apex by the Lorentz forces is aided by the plasma pressure gradient force and is therefore, in the absence of gravity, unopposed.  The addition of a gravitational force that acts downwards from the apex towards foot points introduces opposition to this acceleration.  For even very slight loop twist, Lorentz forces in this case with gravity are still likely to be large enough for the acceleration to take place and fill the loop (Bellan~2007, private communication), but the manner in which the loop is filled with plasma and the subsequent equilibration will be modified by the gravity.  Without gravity, the equilibrium equation is ${\bf j}\times{\bf B}={\bf\nabla} p$, with ${\bf j}$ and $p$ denoting the electric current density and plasma pressure, so that one can immediately see that ${\bf B}\cdot{\bf\nabla} p=0$ and an isothermal plasma must be evenly distributed along each magnetic flux trajectory.  In the presence of a gravitational force some of which acts along the field direction, the plasma is stratified by gravity and cannot be uniformly distributed along the fields.  As well as its non-uniform distribution, the possibility of a low-$\beta$ plasma to reshape a loop's boundary with a dynamically dominant external magnetic field must also be questioned.

This equilibrium problem is the focus of this paper, which is organized as follows.  The mathematical problem is developed in Section~\ref{mhsprob} and solutions presented in Section~\ref{models}.  We conclude with a discussion in Section~\ref{discussion}.

\section{The magnetohydrostatic problem}
\label{mhsprob}

Consider the static equilibrium model based on the
one-fluid ideal hydromagnetic description, denoting the magnetic field,
plasma pressure and density by ${\bf B}$, $p$ and $\rho$, respectively.
The balance of forces is described by

\begin{equation}
\label{fb}
{1 \over 4 \pi} \left(\nabla \times {\bf B}\right) \times {\bf B}
- \nabla p - \rho g {\bf\hat z} = 0 ,
\end{equation}

\noindent
assuming a uniform local gravity of acceleration $g$ in the $z$  direction, along the axis of the loop.  Then the ideal gas law relates the gas pressure $p$ to the gas density $\rho$

\begin{equation}
\label{ideal}
p = \frac{k_B}{\mu} \rho T ,
\end{equation}

\noindent
where $k_B$ is Boltzmann's constant and $\mu$ is the mean particle mass for a fully ionized (monatomic) 
hydrogen plasma.  The solenoidal condition

\begin{equation}
\label{solenoid}
\nabla \cdot{\bf B} = 0 , 
\end{equation}

\noindent
closes the set of equations to determine $p$, $\rho$, and ${\bf B}$.  To keep the physical problem simple, we 
avoid the complication of a full energy equation by applying the isothermal case $T=T_0$ a constant.

In the 2D case in cylindrical geometry with axisymmetry (${\partial /\partial \theta}=0$), the magnetic field satisfying equation~(\ref{solenoid}) has the form

\begin{equation}
{\bf B}=\frac{B_0}{r}\left( -\frac{\partial\psi (r,z)}{\partial z} ,f (r,z),  \frac{\partial\psi (r,z)}{\partial r}\right) ,
\label{Bfield}
\end{equation}

\noindent
 where $f(r,z)/r$ is the toroidal magnetic field component and $\psi (r,z)$ is the magnetic flux function, whose isosurfaces contain the magnetic field trajectories.  By standard theory (Low 1975), the momentum equation~(\ref{fb}) reduces to

\begin{eqnarray}
\frac{\partial^2\psi}{\partial r^2} -\frac{1}{r}\frac{\partial\psi}{\partial r} +\frac{\partial^2\psi}{\partial z^2} +f(\psi )\frac{df(\psi )}{d\psi} +4\pi\frac{\partial p(\psi , z)}{\partial\psi} & = & 0,\label{reducedfb}\\
\left.{\frac{\partial p(\psi, z)}{\partial z}}\right|_{\psi =\rm{const}} +\rho (\psi ,z) g(z) & = & 0\label{hydrostatic} .
\end{eqnarray}

\noindent
Here $f(r,z)=f(\psi )$ must be a strict function of $\psi$ from the projection of the momentum equation on the symmetry axis.

 \section{The models}
 \label{models}
 
 This section focuses on the static equilibria before and after a loop is filled with plasma.  This paper does not directly treat the dynamics of the evolution to a final equilibrium, but it is shown that the final equilibrium can be reached from the initial equilibrium via ideal MHD processes.  The initial equilibrium model describes an empty loop, i.e. $p=\rho =0$, and so there are no forces in this equilibrium.  The initial equilibrium is therefore described by a force-free field.  The final equilibrium does involve forces and requires a magnetohydrostatic (MHS) description.  We describe linear force-free models for the initial equilibrium in Subsection~\ref{lff} and related linear MHS equilibria in Subsection~\ref{mhs}.
 
\subsection{Linear force-free fields}
\label{lff}

The force-free case is the one with $p_0 (\psi ) =$~constant (zero in our scenario) so that all Maxwell stresses are contained within the magnetic field and all electric currents are aligned with the magnetic field.  The current associated with the magnetic field of equation~(\ref{Bfield}) is described by

\begin{equation}
{\bf j}=B_0\left( -\nabla^2\psi , \frac{d f(\psi)}{d\psi}\frac{\partial\psi}{\partial z} , -\frac{d f(\psi)}{d\psi}\frac{\partial\psi}{\partial y} \right) ,
\end{equation}

\noindent
The general condition for the nonlinear field to be force-free is that equation~(\ref{reducedfb}) is satisfied with $\partial p(r,z)/\partial z = 0$, in which case ${\bf j}=\alpha (r,z){\bf B}$ with the nonlinear force-free parameter $\alpha (r,z)=df(\psi )/d\psi $ - see equation~(\ref{reducedfb}).  The field-aligned currents are clearly associated with axial magnetic flux $B_{\theta}$ which is responsible for twist and shear in the magnetic field in these 2D solutions.  For a force-free field to be solenoidal, $\alpha$ must be constant along each flux trajectory: ${\bf B}\cdot{\bf\nabla}\alpha =0$.  The simplest case is the potential case $f(\psi )=\alpha =0$ which has no electric currents, no azimuthal field component and therefore no twist.  Of interest here is the linear case $f(\psi )=\alpha\psi$, for some constant $\alpha$, which is seen from equation~(\ref{reducedfb}) to yield the usual Helmholtz equation for linear force-free fields $(\nabla^2 +\alpha^2)\psi =0$.  For a narrow equilibrium structure like a loop, this linear case should be sufficient.

\begin{figure*}[ht]
\begin{center}
\resizebox{0.40\hsize}{!}{\includegraphics*{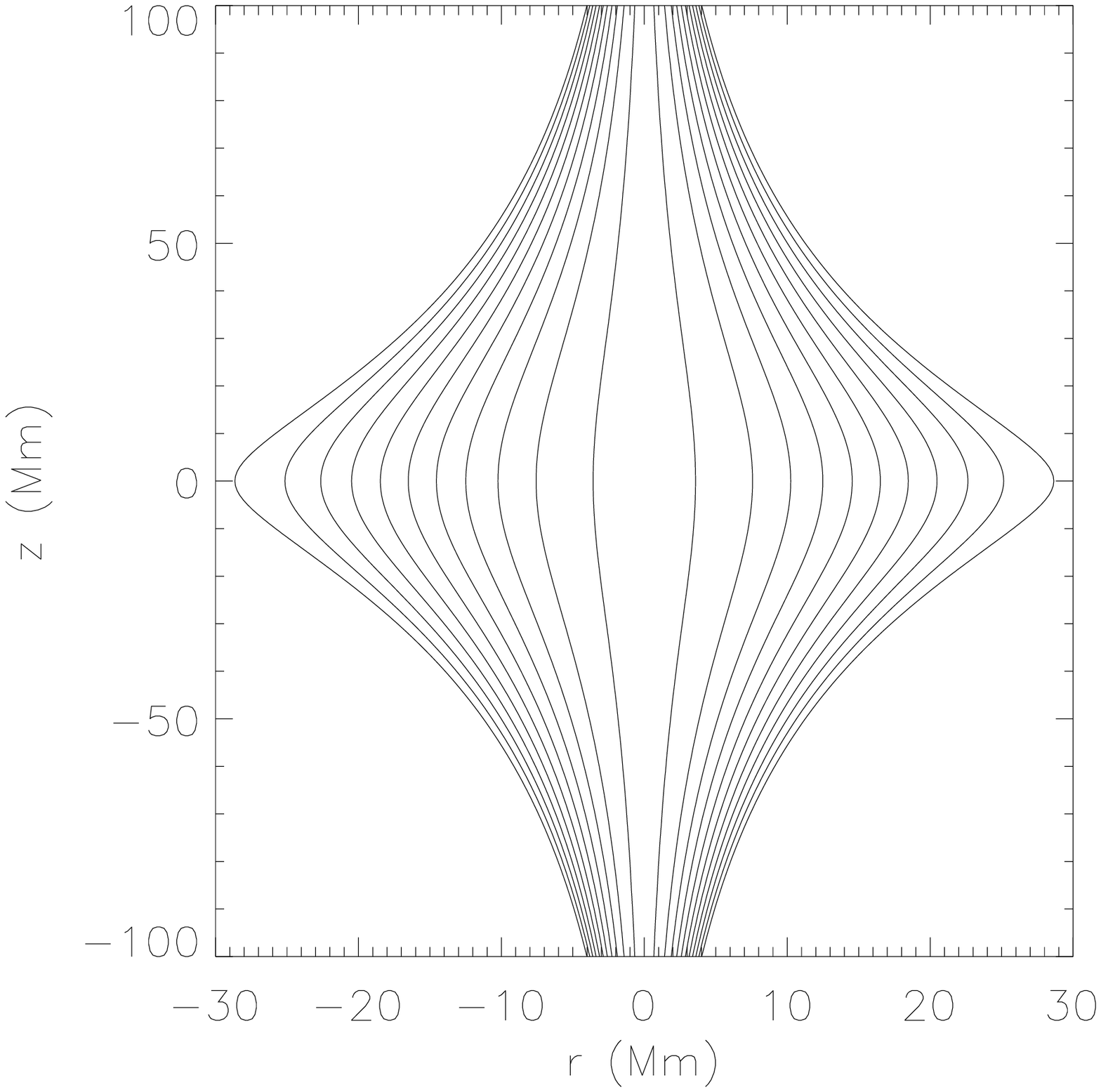}}
\resizebox{0.40\hsize}{!}{\includegraphics*{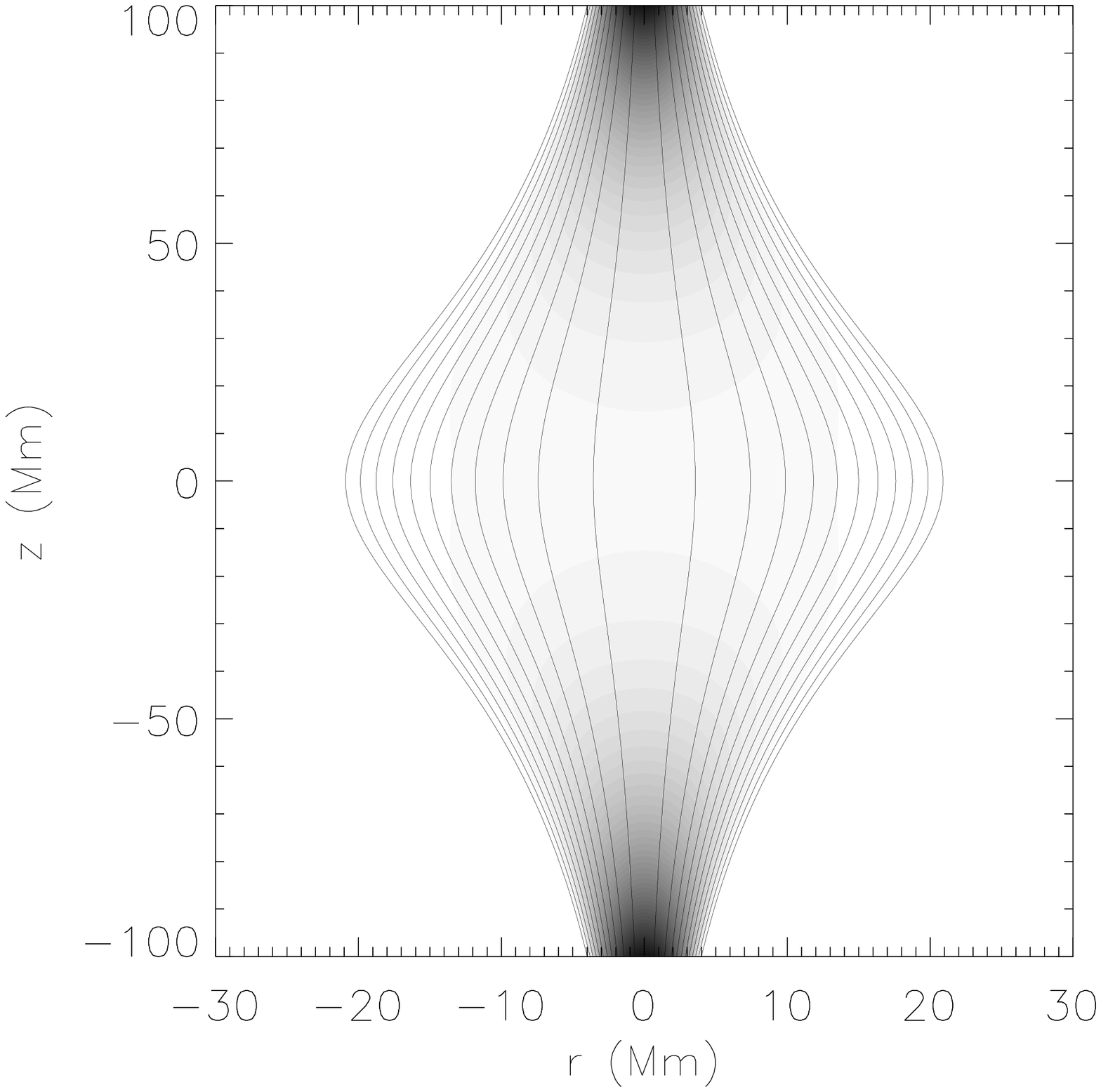}}
\resizebox{0.40\hsize}{!}{\includegraphics*{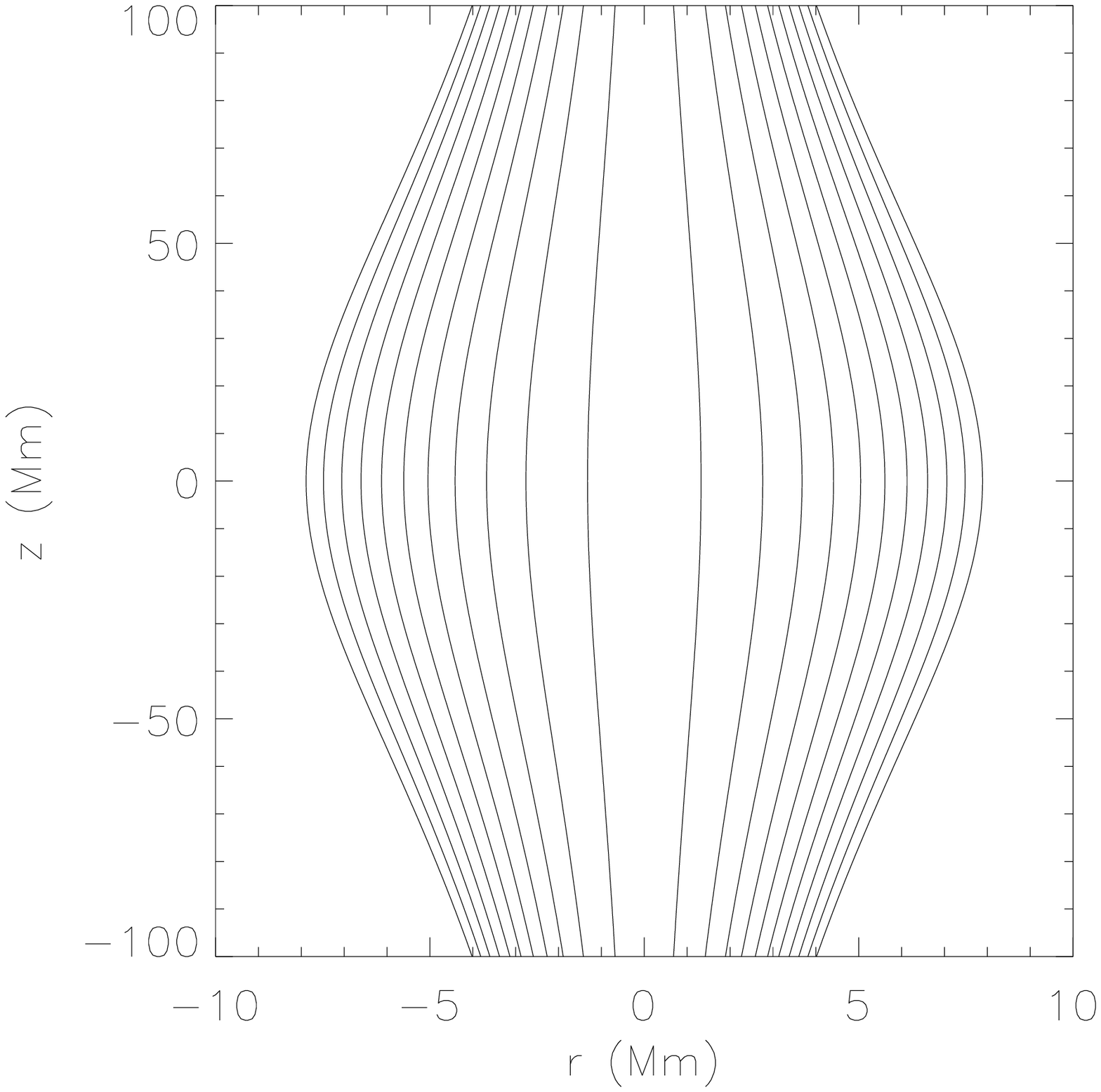}}
\resizebox{0.40\hsize}{!}{\includegraphics*{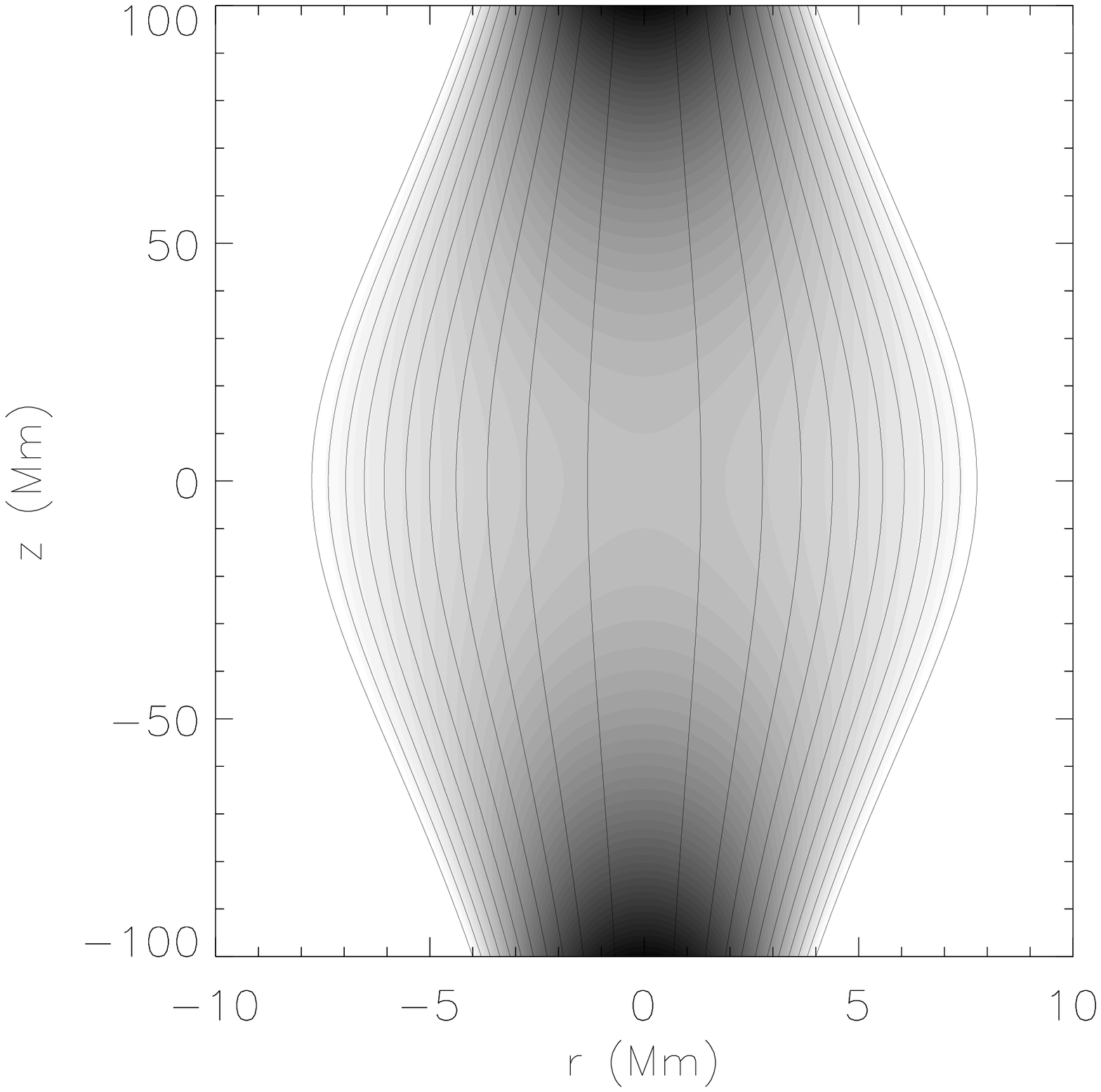}}
\resizebox{0.40\hsize}{!}{\includegraphics*{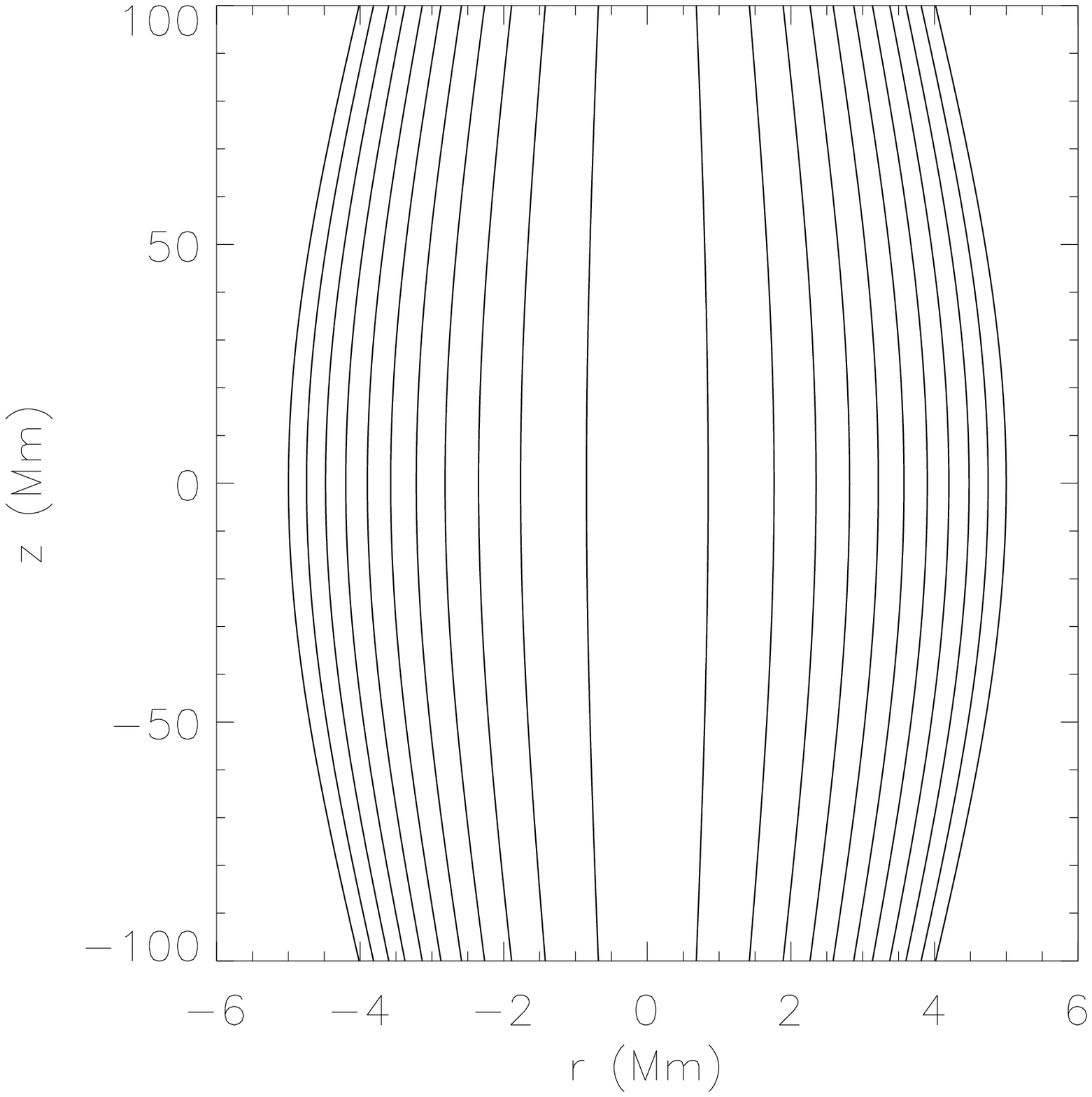}}
\resizebox{0.40\hsize}{!}{\includegraphics*{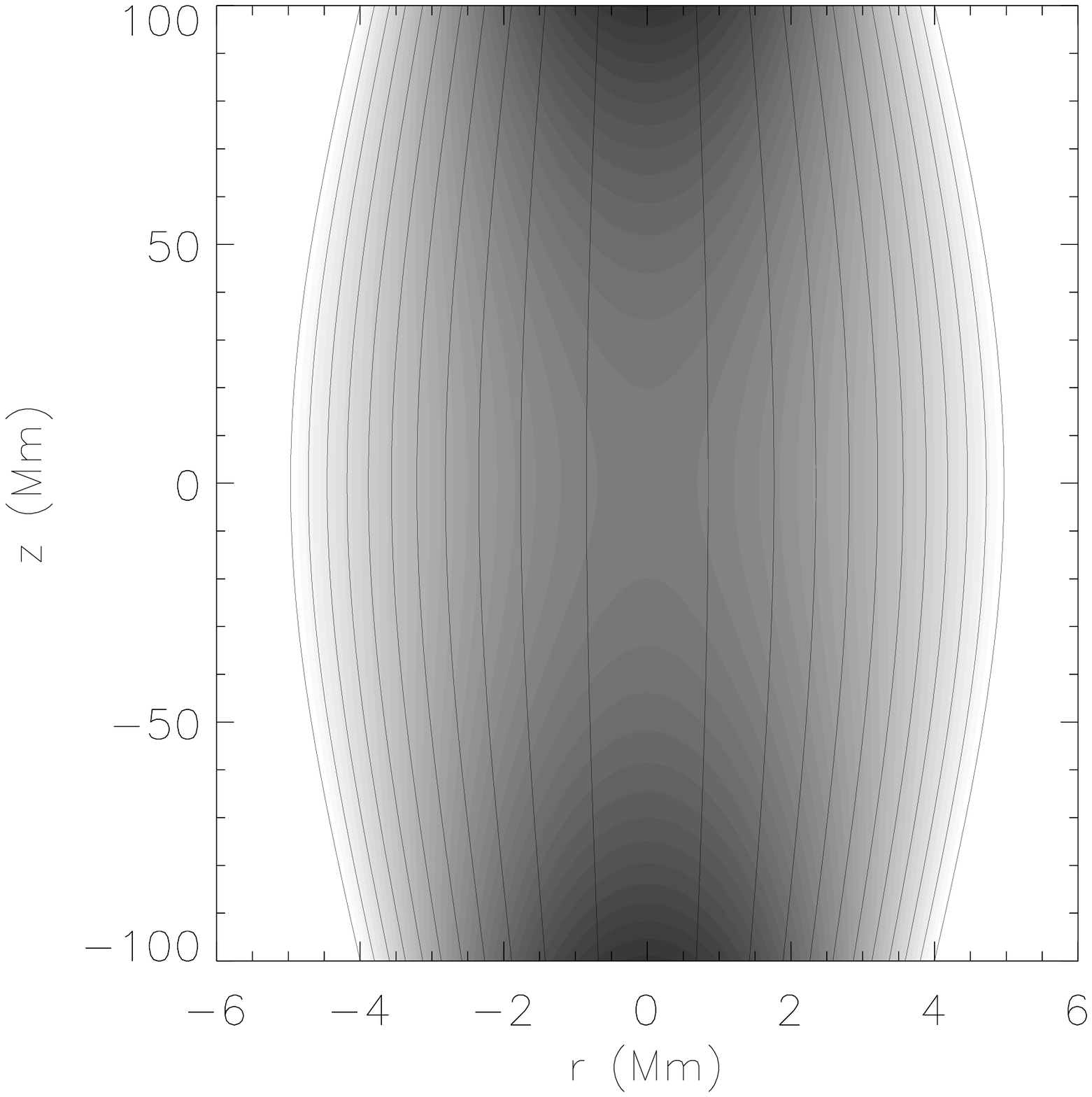}}
\end{center}
\caption{Linear force-free solutions (left pictures) and magnetohydrostatic (MHS) solutions including forces (right pictures) corresponding to stratification length scales ($1/m$ for force-free solutions, $\Lambda$ for MHS solutions) of 25~Mm (top), 50~Mm (middle) and 100~Mm (bottom).  All solutions plotted here have $\alpha=1/16$~Mm$^{-1}$.}
\label{coshloops}
\end{figure*}

\begin{figure*}[ht]
\begin{center}
\resizebox{0.49\hsize}{!}{\includegraphics*{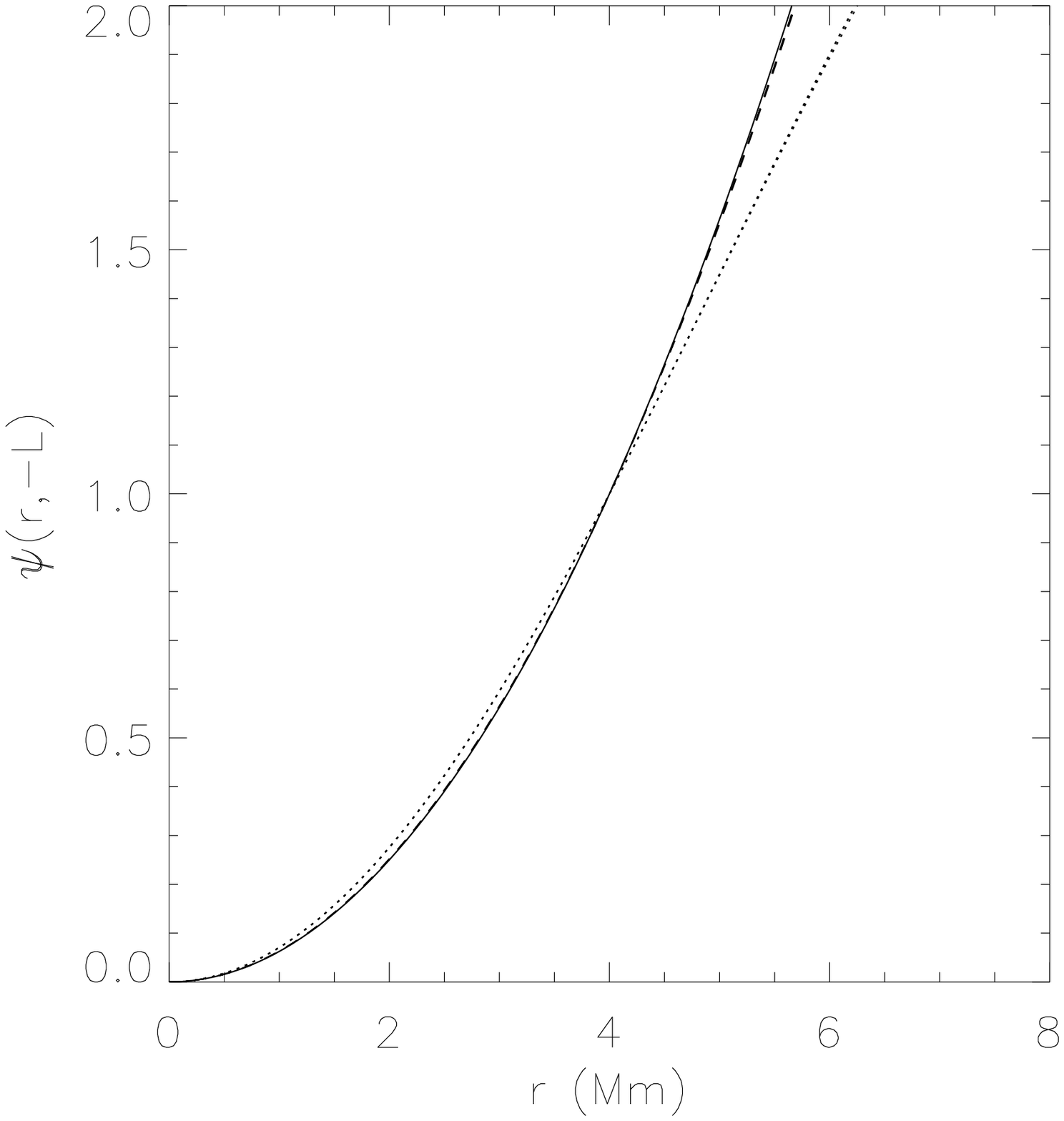}}
\resizebox{0.49\hsize}{!}{\includegraphics*{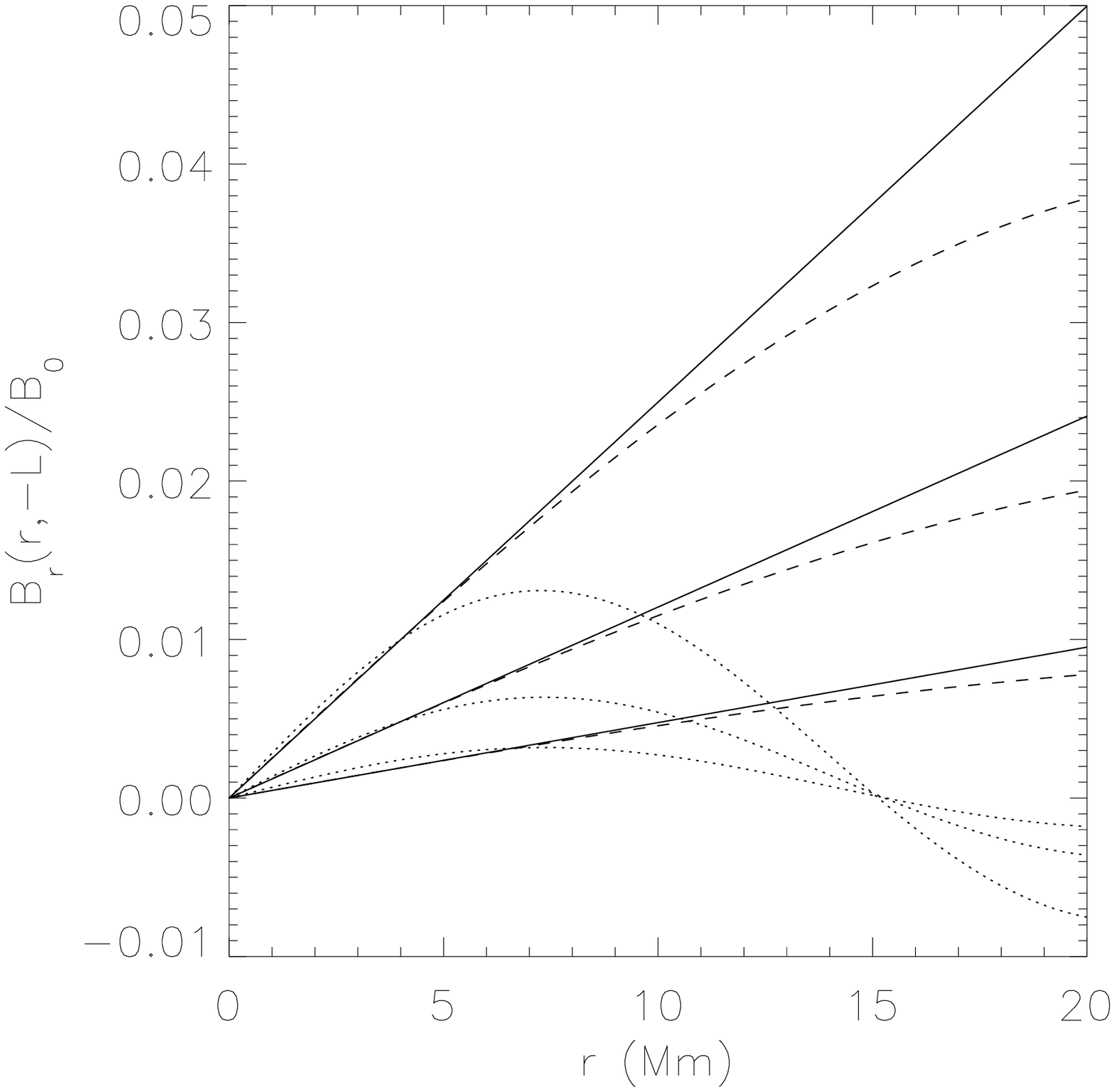}}
\resizebox{0.49\hsize}{!}{\includegraphics*{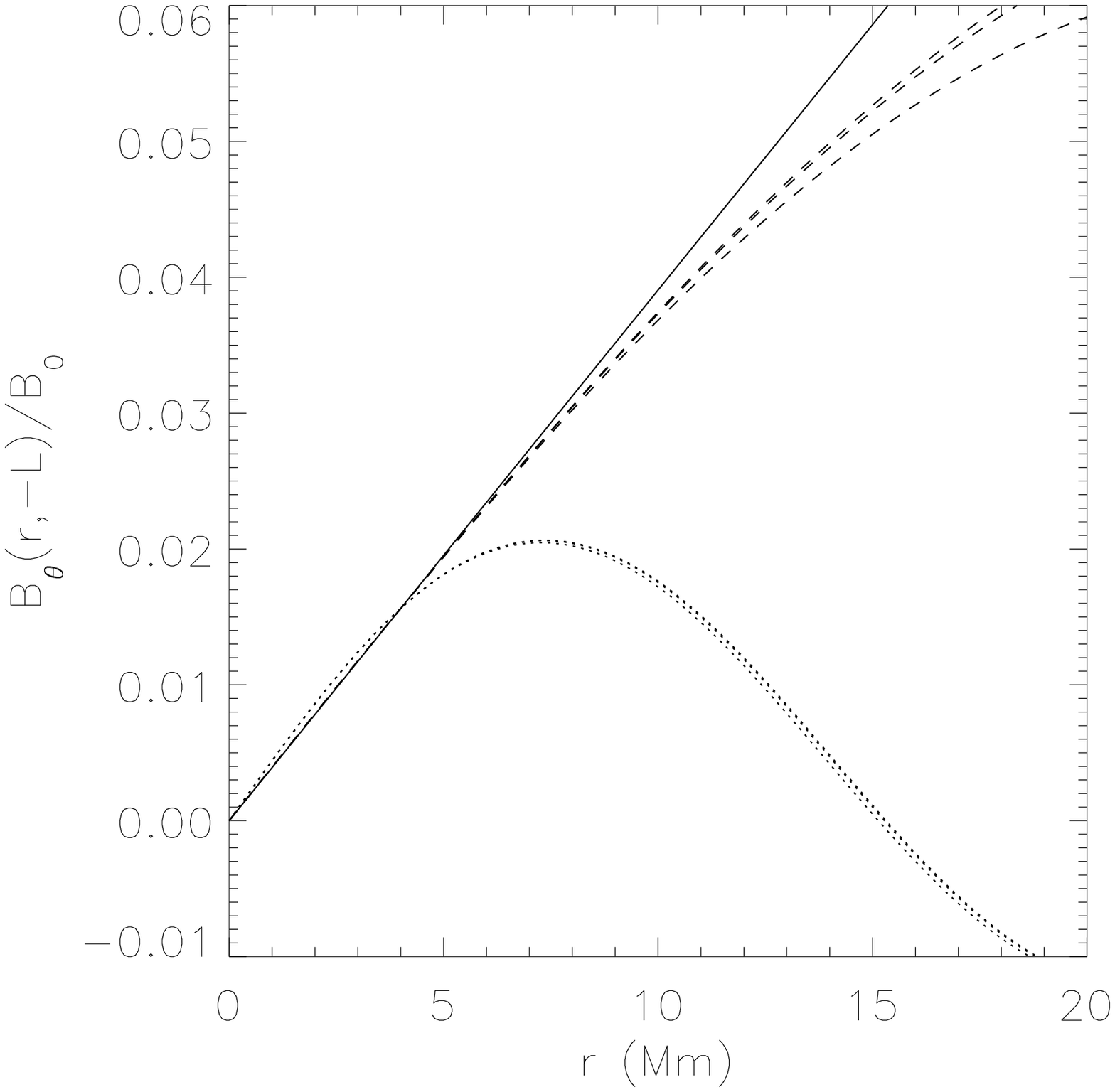}}
\resizebox{0.49\hsize}{!}{\includegraphics*{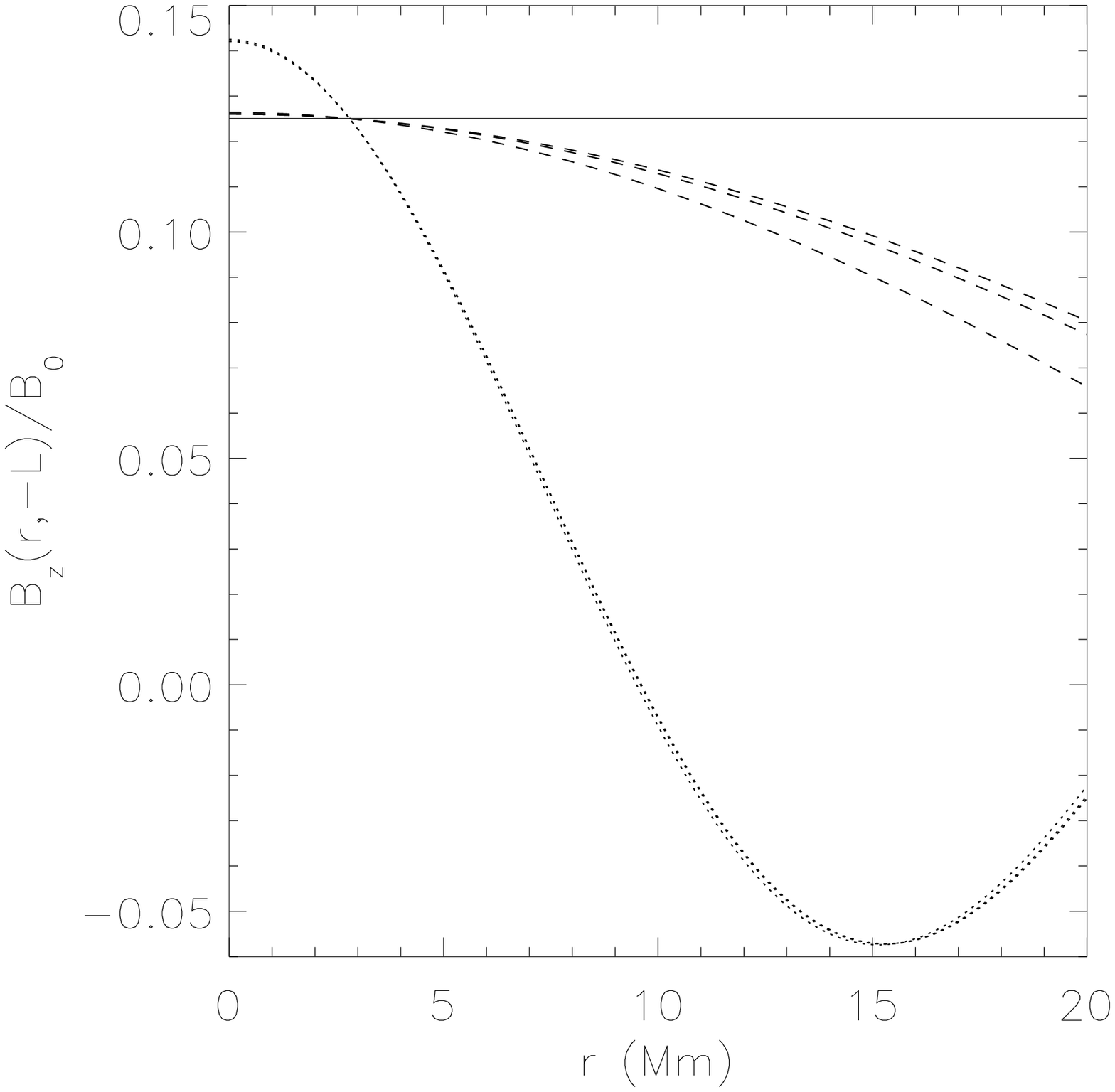}}
\end{center}
\caption{A direct comparison between the flux function (top left) and magnetic field component distributions (the three other pictures) at the boundary $z=-L$ of the MHS solution (solid lines) with various cases of the force-free solution: $k=(\alpha^2 +m^2)^{1/2}$ for $\alpha =1/16$ (dashed lines) and $\alpha =1/4$ (dotted lines).  The force-free and MHS solutions are plotted for the cases $m=1/\Lambda =1/25$~Mm$^{-1}$, $1/50$~Mm$^{-1}$ and $1/100$~Mm$^{-1}$.  In the top right picture, these cases are distinct because of the proportionality of $B_r$ to $m$.  In the other pictures, the force-free cases coincide to a good approximation, while the MHS cases are exactly equal .  For comparison, all force-free solutions plotted in Figure~\ref{coshloops} have $\alpha =1/16$~Mm$^{-1}$.}
\label{compare}
\end{figure*}

Browning \& Priest~(1983) found three linear force-free solutions of equation~(\ref{fb}), one of which lends itself to the modeling of an expanding coronal loop:

\begin{eqnarray}
\psi_{ff} (r,z) & = & \psi_0 rJ_1 (kr) \cosh (mz),\ \ \ m^2=k^2-\alpha^2,\label{coshsol}
\end{eqnarray}

\noindent
where $\alpha$ is constant (henceforth we will only treat constant-$\alpha$ cases).  Three examples of the solution described by equation~(\ref{coshsol}) are plotted in Figure~\ref{coshloops}, for different values of $m$ but all with $\alpha =1/16$.  We expect that during dynamical phases, the loop will remain anchored to the base of the atmosphere since the photospheric plasma is orders of magnitude denser than the coronal plasma, and should not be affected by coronal field evolution of this type.  The footprint of the loop at the base of the corona $z=\pm 100$~Mm is fixed to be the circle $r\le a$~Mm by setting 

\begin{equation}
\psi_0 = \frac{1}{a J_1(ka)\cosh (mL)} ,\label{psiff}
\end{equation}

\noindent
where $L=100$~Mm is the loop half-length, and by adopting the surface $\psi =1$ as the loop boundary.  Here we show examples with $a=4$~Mm,which corresponds reasonably to observations by Aschwanden et al.~(1999) and Watko \& Klimchuk~(2000) and will allow other physical parameters to take reasonable values as described later.  In Figure~\ref{compare} the top left picture shows $\psi |_{z=\pm L}$: the cases $k=(\alpha^2 +m^2)^{1/2}$ for $\alpha =1/16$ are graphed as the dashed lines, all almost coinciding with the solid line, and $\alpha =1/4$ (dotted lines).  Each of these distributions is shown for three values of $m$: $m=1/25$~Mm, $m=1/50$~Mm and $m=1/100$~Mm, corresponding to the left pictures of Figure~\ref{coshloops}.  However, the differences between these different values for $m$ are small in the graphs because $\alpha$ is more dominant than $m$ in the expression for $k$: for the length scales of interest, the twist affects the boundary flux distribution more than the axial variations do.  For small $k$, $\psi_{z=\pm L}$ is very close to the solid line, which describes the parabola $\psi =r^2$.  This is the flux function (at $z=\pm L$) of the related MHS equilibrium, which we introduce in the next subsection. The magnetic field vector is

\begin{equation}
{\bf B}_{ff} = B_0\left[ m J_1 (kr)\sinh (mz) , \alpha J_1 (kr) \cosh (mz) , kJ_0 (kr) \cosh (mz)\right] /
\left[a J_1(ka)\cosh(Lm)\right].\label{Bff}
\end{equation}

The inverse length scales $k$, $\alpha$ and $m$ all influence the character of the solution in distinctive ways.  The inverse length scale $m$ controls the shape of the loop: increasing $m$ gives the loop a more of a bulging cross-section because $B_r$ is proportional to $m$.  Figure~\ref{compare}, (top right) shows example distributions of $B_r|_{z=-L}$ for selected cases of the force-free solution: $k=(\alpha^2 +m^2)^{1/2}$ for $\alpha =1/16$ (dashed lines) and $\alpha =1/4$ (dotted lines).  These graphs are shown for three values of $m$: $m=1/25$~Mm, $m=1/50$~Mm and $m=1/100$~Mm, corresponding to the left pictures of Figure~\ref{coshloops}.  The equivalent graphs for $B_r |_{z=+L}$ are the mirror image of these in the $r$-axis.  The correspondence between the increasing loop bulge and the increasing $B_r$ as $m$ increases is clear.  The $\alpha$ parameter, when non-zero, adds twist to the magnetic field, since $B_{\theta}$ is directly proportional to $\alpha$.  The bottom left picture of Figure~\ref{compare} shows example distributions of $B_{\theta}|_{z=\pm L}$ with $\alpha =1/16$, the value of all force-free examples in Figure~\ref{coshloops}.  Shown are cases with $k=(\alpha^2 +m^2)^{1/2}$ for $\alpha =1/16$ (dashed lines) and $\alpha =1/4$ (dotted lines) for comparison.  The case $\alpha =1/16$ is representative of all examples in Figure~\ref{coshloops} (left pictures) since the dashed lines are close together, particularly within the loop.  The axial flux $B_z$ is proportional to the parameter $k$.  Example distributions of $B_z|_{z=\pm L}$ are shown in the bottom right picture of Figure~\ref{compare}, again for $k=(\alpha^2 +m^2)^{1/2}$ with $\alpha =1/16$ (dashed lines) and $\alpha =1/4$ (dotted lines).  The dashed lines are close to each other, and correspond to all three pictures in the left half of Figure~\ref{coshloops}.  The distributions of $B_z|_{z=\pm L}$ and $B_{\theta}|_{z=\pm L}$ are effectively independent of $m$ while the distribution of $B_r|_{z=\pm L}$ is not.  This is because $B_r$ is related to the bulge of the loop to a significant degree and $B_{\theta}$ and $B_z$ are not.

The three parameters $k$, $\alpha$ and $m$ are not independent of each other: the toroidal flux is limited by the axial flux strength via the condition $m^2=k^2-\alpha^2\ge 0$ which then fixes the radial field strength.  A 1D configuration is possible if the toroidal field strength provides exactly strong enough magnetic tension  to balance the expansive magnetic pressure gradient force of the axial flux, when $\alpha =k$.  This is a limiting case beyond which the toroidal flux is too strong for an equilibrium solution to exist.  Increasing $k$ increases the expansion of the flux tube while increasing $\alpha $ draws in this expansion.  The parameter $k$, and these $\psi$ and $\bf B$ distributions generally, will receive further attention after we have introduced the MHS solutions in the next subsection: for small $k$, the field component distributions are strikingly similar to the related MHS solutions that form the subject of that subsection.  In practice, length scales of variation along loops are so large that the inverse length scale $m$ is generally very small, and interesting cases have $k\approx\alpha$ and $\alpha >0$.  At least as far as this solution class is concerned, electric currents are necessary for interesting physics.  Bellan~(2003) needed a toroidal field component to be present for his process to work.

We omit here the effects of a curved loop axis, assuming that the effects of curvature are not important to the physics under consideration.  This assumption has been applied many times in the study of twisted loop equilibria in cases with or without forces (Parker 1979, 1994, Browning \& Priest 1983, Zweibel \& Boozer 1985, Browning \& Hood 1989, Lothian \& Hood 1989, Parker~1994, Beli\"en et al. 1997a, b, Van der Linden \& Hood 1999, Bellan 2003).  The current self-repulsion or hoop force characteristic of curved current structures is unlikely to be important in for the physics of a typical loop: in the scenario described by Gold~(1964) and Parker~(1994), a loop is one modestly twisted flux tube among many flux tubes, probably separated by current sheets in general, and a loop's self-repulsion is likely to be much weaker than local pressure balance between neighboring flux tubes which will squeeze the loop or else allow it space in which to expand.  The force-free solutions of the previous section demonstrate the tendency of force-free fields to expand to fill an atmosphere, unless reined in by the tension force of strong azimuthal flux.  Our objective here is to determine whether or not plasma forces increase the likelihood of a slightly twisted flux tube assuming a constant cross-sectional shape in a gravitationally stratified medium, as Bellan~(2003) found in the case without gravity.

\clearpage

\begin{figure*}[ht]
\begin{center}
\resizebox{0.69\hsize}{!}{\includegraphics*{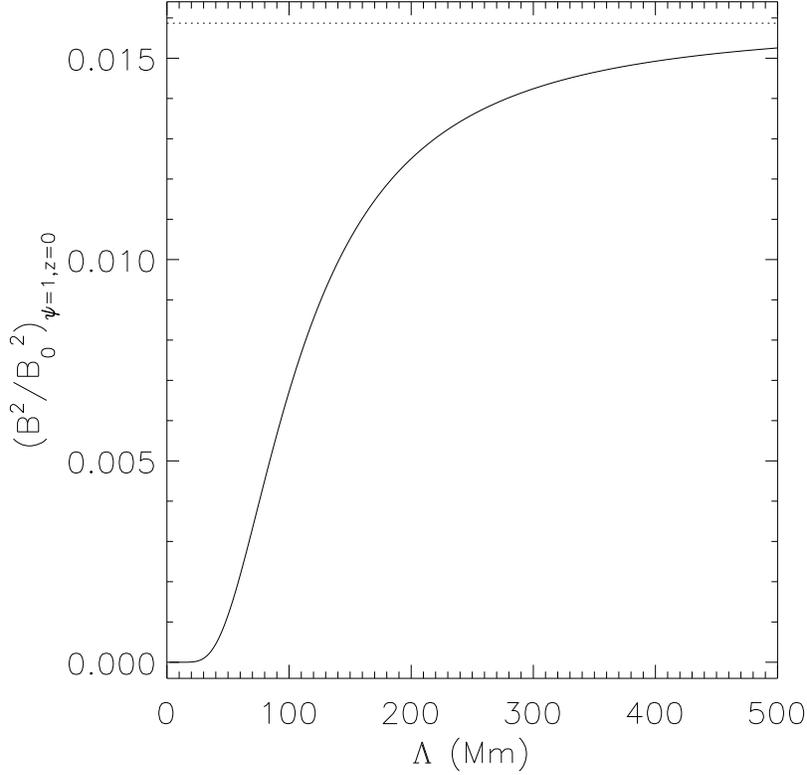}}
\end{center}
\caption{Shown by the dotted line is the magnetic pressure (and therefore total pressure) on the loop boundary $\psi =1$ at the apex $z=0$ as a function of hydrostatic scale height $\Lambda$ for our special class of MHS equilibria.  This apex boundary pressure increases as $\Lambda$ increases and the loop solution becomes narrower.  As $\Lambda\rightarrow\infty$ this pressure tends to a constant, represented by the dotted line.  This is the boundary pressure of Bellan's~(2003) solution, the narrowest possible solution.  The total boundary pressures for the equivalent force-free solutions with $m=1/\Lambda$ are smaller.  Therefore, the much enhanced boundary pressure for large $\Lambda$ makes it impossible for a loop to evolve from an expanded force-free configuration to a non-force-free configuration with significantly reduced expansion.}
\label{strengths}
\end{figure*}

\clearpage

\subsection{Inclusion of forces associated with isothermal plasma}
\label{mhs}

We now turn our attention to non-force-free MHS solutions.  Equation~(\ref{hydrostatic}) is solved by

\begin{equation}
p(\psi ,z)=p_1(\psi )-\int_0^z \rho (\psi ,z')g(z) dz'.
\label{pintegral}
\end{equation}

\noindent
In general, the Grad-Shafranov equation~(\ref{reducedfb}) can assume as many different forms as there are distinct regions of physical space, each of which may be occupied by a different branch of the functions $p(\psi ,z)$ and $\rho (\psi ,z)$.  The distinct regions are separated by free boundaries, the loci of which are unknown and generally can only be calculated simultaneously with the solution $\psi$.  The free boundaries are governed by continuity or jump conditions imposed by force balance at these boundaries.  Free boundaries also separate regions where the functional forms of $p(\psi ,z)$ and $\rho (\psi ,z)$ differ, e.g., a region of potential or force-free magnetic field separated from a region of full magnetohydrostatic force balance.  The nonlinear equations involved require a full numerical treatment along the lines of Petrie et al.~(2007).  We defer such a study and choose to stay close to the linear treatments of past authors, including Bellan~(2003), to enable a direct comparison.

For simplicity we set $f^2 (\psi )=A\psi$, so that equation~(\ref{reducedfb}) is linear.  In the isothermal case, the hydrostatic scale height $\Lambda =p/(\rho g)$ is constant.  Now the condition $p/\rho =$~constant gives $p_1(\psi )=0$ in equation~(\ref{pintegral}).

We seek only separable forms for $p(\psi ,z)$:

\begin{equation}
p(\psi ,z)=p_0(\psi )s(z).
\end{equation}

\noindent
The Grad-Shafranov equation~(\ref{reducedfb}) becomes

\begin{equation}
\frac{\partial^2\psi}{\partial r^2} -\frac{1}{r}\frac{\partial\psi}{\partial r} +\frac{\partial^2\psi}{\partial z^2} +f(\psi )\frac{df(\psi )}{d\psi} +4\pi r^2\frac{d p_0(\psi)}{d \psi}s(z) =0 .
\label{nonsepgs}
\end{equation}

\noindent
Equation~(\ref{nonsepgs}) generally needs to be solved numerically (Beli\"en~1997a, b).  However, if we prescribe $g(z)$ using an elementary analytical function:

\begin{equation}
g(z)=g_0\tanh \left(\frac{z}{\Lambda}\right), \label{gform}
\end{equation}

\noindent
where $g_0$ is the solar surface gravitational acceleration 274~m/s$^2$, then $g(z)$ models the effect of gravity acting in opposite directions along each leg of the loop, in the negative direction in the left leg ($z<0$) and in the positive direction in the right ($z>0$) leg.  This form of $g(z)$ tends to the uniform surface gravitational acceleration at the loop foot points $z=\pm 1$, where the field is assumed to be nearly vertical, and changes sign at the apex of the loop $z=0$ where the loop is locally horizontal.  This $g(z)$ therefore captures the global effect of gravitational stratification on a twisted loop while ignoring the component of gravity perpendicular to the curved loop axis.  This is reasonable for the loops of large aspect ratio (length/width) that concern us here, since the effects of axis curvature are not large for such loops (Beli\"en et al. 1997a, b).

Now equations~(\ref{pintegral},\ref{gform}) together determine the dependence on $z$ of $p$.  With $g(z)$ given by equation~(\ref{gform}), $s(z)$ must take the form

\begin{equation}
s(z)=\cosh \left(\frac{z}{\Lambda}\right),
\end{equation}

\noindent
which, in the near-vertical portion of the loop near the foot points models the exponential fall-off of pressure and density characteristic of hydrostatic gravitational stratification.  The pressure and density are both minimized at the apex $z=0$.  A further advantage of this form for $g(z)$ is that equation~(\ref{nonsepgs}) becomes separable such that exact analytical solutions can be obtained, and the resulting MHS solution can be easily compared to the linear force-free field solution of equation~(\ref{coshsol}) with $m=1/\Lambda$.
Defining $\psi =\psi_1$ as the flux surface on which $p$ vanishes and scaling $\psi$ so that $\psi =1$ on this surface, if we assume also a linear form for $p_0(\psi )$,

\begin{equation}
p_0(\psi ) =P_0(1-\psi ), \label{plinear}
\end{equation}

\noindent
then we have reduced the problem to a linear equation:

\begin{equation}
\frac{\partial^2\psi}{\partial r^2} -\frac{1}{r}\frac{\partial\psi}{\partial r}+\frac{\partial^2\psi}{\partial z^2}+\alpha^2\psi-4\pi P_0r^2=0,\label{lineargs}
\end{equation}

We seek a particular integral of the Grad-Shafranov equation~(\ref{lineargs}) that balances the source term $4\pi P_0r^2$.  If we fix $\psi =1$ on the flux surface passing through $r=a$ at $z=\pm L$ where $L=100$~Mm is the loop half-length, then

\begin{equation}
4\pi P_0=B_0^2\frac{1/\Lambda^2+\alpha^2}{a^2 \cosh (L/\Lambda )} ,
\end{equation}

\noindent
where $L=100$~Mm is the loop half-length.  This particular integral is therefore

\begin{equation}
\psi_{mhs}=r^2\cosh \left(\frac{z}{\Lambda}\right)\left/  a^2 \cosh \left(\frac{L}{\Lambda}\right)\right. ,\label{partint}
\end{equation}

\noindent
and the magnetic vector is

\begin{equation}
{\bf B}_{mhs} = B_0\left[ \frac{r}{\Lambda} \sinh \left(\frac{z}{\Lambda}\right) , \alpha r \cosh \left(\frac{z}{\Lambda}\right) , 2 \cosh \left(\frac{z}{\Lambda}\right) \right] \left/
a^2 \cosh \left(\frac{L}{\Lambda}\right) \right. .\label{Bmhs}
\end{equation}

\noindent
We plot example solutions in Figure~\ref{coshloops} (right pictures).  The effect of gravitational stratification is clear: the plasma is concentrated close to the foot points of the loop and the loop is not of uniform cross-sectional width.  We have chosen here cases with hydrostatic scale height $\Lambda =25$, $50$ and $100$~Mm corresponding to temperatures of about $0.5$, $1$ and $2$~MK, respectively.  In the limiting case with $\Lambda\rightarrow\infty$, corresponding to a very hot plasma (or a scaled-down loop) we recover the axially uniform case found by Bellan~(2003) in which gravity does not play a significant role.

The plasma $\beta$ on the loop axis is given by

\begin{equation}
\beta|_{r=0} = \left(\frac{8\pi p}{B_z^2}\right)_{r=0} =\frac{a^2}{2}\left(\frac{1}{\Lambda^2} +\alpha^2\right)\frac{\cosh (z/\Lambda )}{\cosh (L/\Lambda)} ,
\end{equation}

\noindent
which at the boundaries $z=\pm L$ is equal to $(1/\Lambda^2 +\alpha^2 )/2\approx 0.044$, $0.034$ and $0.032$ in the three cases $\Lambda =L$, $L/2$ and $L/4$, respectively in Figure~\ref{coshloops} with $a=4$~Mm.  At the apex $z=0$, $\beta$ is a factor of $\cosh (L/\Lambda )$ smaller, and falls to about 0.021, 0.0092 and 0.0016 for the three cases $\Lambda =L$, $L/2$ and $L/4$, respectively.  Obviously these $\beta$ values decrease with distance from the loop axis $r=0$.

The angle between a field line and the axial direction at any point, the pitch angle, is $\phi$ where $\tan (\phi ) = B_{\theta} /B_z$.  A field line makes one complete revolution around the loop axis in a distance

\begin{equation}
\lambda (r,\theta ) = 2\pi r\frac{B_z}{B_{\theta }} =\frac{2\pi r}{\tan\phi } .
\end{equation}

\noindent
Thus

\[ \lambda = \left\{ \begin{array}{ll} 2\pi r k J_0 (kr)/\alpha J_1 (kr) , & \mbox{force-free case,} \\ \
4\pi /\alpha , & \mbox{MHS case.} \end{array} \right. \]

\noindent
In all of our examples, $\alpha =1/16$~Mm$^{-1}$, and so in our MHS solutions $\lambda =64\pi $~Mm and the lines turns through nearly $2\pi$ radians along the entire length of each loop.  This is within the Kruskal-Shafranov limit of $2\pi$ turns for a simple toroidal configuration (Kruskal~1954, Shafranov~1957), often considered a representative condition for stability against kink modes (e.g. Priest~1982, Goedbloed \& Poedts~2004).  In the force-free models considered here, with $k\approx\alpha =1/16$~Mm$^{-1}$, the result is similar.  The stability properties of these solutions can be determined in the form of a full resistive MHD spectrum by solving the linearized MHD equations using the hyperbolic stability solver PHOENIX (Blokland et al.~2007).  This will be the subject of a later study.

This set of non-force-free MHS solutions does not include the force-free solution of equation~(\ref{coshsol}) as a special case: setting $P_0=0$ in equation~(\ref{partint}), implying that $1/\Lambda^2+\alpha^2=0$, just gives the trivial solution $\psi\equiv 0$.  However, the two solutions can still be related to each other in a meaningful way.  Compared to this non-force-free solution, the force-free solution's photospheric axial flux distribution is more concentrated towards the central axis. For small $k$ the force-free solution is very close to the non-force-free solution but, since $k=0$ forces $m=\alpha =0$, $k>0$ for any non-trivial solution.  The resemblance between the force-free and non-force-free solutions plotted in Figure~\ref{coshloops} is due to the small size of $k$.  Since axial length scales in the solar corona are so large, measured in tens of Mm, this means that $m^2$ is small and so $k$ and $\alpha$ are of similar size in the examples plotted in Figure~\ref{coshloops}.  In the top left picture Figure~\ref{compare} the magnetic flux functions $\psi_{ff}$ and $\psi_{mhs}$, given by Equations~(\ref{psiff}) and (\ref{partint}),are graphed for comparison of the non-force-free and force-free solutions.  While these graphs show photospheric distributions, multiplying by $\cosh (mz)$ or $\cosh (z/\Lambda )$ gives the force-free or MHS distribution for any $-L<z<L$.  For $k<1$, $J_1(kr)$ is nearly linear in the domain of interest, and so $\psi_{ff}$ and $\psi_{mhs}$ are nearly equal.  Since $J_1$ has positive second derivative in this domain of interest, $\psi_{ff}$ is slightly larger than $\psi_{mhs}$ everywhere between $r=0$ and $r=a$ except on the axis $r=0$ where both are zero, and at the outer boundary where both solutions are $1$.  This implies that evolving from $\psi_{ff}$ to $\psi_{mhs}$ would involve a small outward movement of each flux surface $\psi =$~constant wherever $r<a$, as expected when gas is pumped into a flux tube adding to the outward pressure force.  On the other hand, for $r>a$, $\psi_{ff}$ is smaller than $\psi_{mhs}$, as indicated by Figure~\ref{compare}, meaning that all flux surfaces outside the cylinder $r=a$ actually move inward during evolution from $\psi_{ff}$ to $\psi_{mhs}$. This is because there is less plasma out here than close to the axis, so that outward pressure forces are not greatly enhanced here, and the outward displacement of axial magnetic flux by the plasma causes the axial field component to be stronger than in the force-free case, without enhancing the radial component, so that the flux tube has less bulge.  This effect is greater for greater $m$ and smaller $\Lambda$, in which case more of the tube lies outside the cylinder $r=a$ and is therefore subject to this inward movement.  This difference between empty and full flux tubes is therefore reminiscent of the picture presented by Bellan~(2003) of plasma injection causing a flux tube to become thinner, although the change is much smaller than in the case focused on by Bellan.

All of the MHS examples of Figure~\ref{coshloops} (right pictures) have $B_z|_{z=\pm L} =B_0/8$~G and $B_{\theta} =B_0r/256 $~G.  These are the solid curves in the bottom pictures of Figure~\ref{compare}.  Within the loop the boundary distributions of the axial and azimuthal field components $B_z|_{z=\pm L}$ and $B_{\theta}|_{z=\pm L}$ are very similar for the force-free models represented by the dashed curve in each picture, and the MHS models represented by the solid curve.  The force-free axial field distribution decreases very slowly with radius, is larger than the MHS value at the center of the loop and crosses the MHS value at $r\approx 0.7 a$~Mm.  Meanwhile the azimuthal component is slightly greater than the MHS component for $0<r<a$, matching it at $r=0,a$.

Unlike $B_{\theta} |_{z=\pm L}$ and $B_z |_{z=\pm L}$, $B_r |_{z=\pm L}$ is strongly affected by changes in $m$ or $\Lambda $.  Figure~\ref{compare} (top right picture) shows the radial field distributions at $z=-L$ for the three values of $1/m$ and $\Lambda $ represented in Figure~\ref{coshloops}: 25~Mm, 50Mm and 100~Mm.  $B_r$ is proportional to this parameter and so the greater the bulge of the loop, the greater the radial field component.  Within the loop the MHS radial field of a given value of $\Lambda =\Lambda_0$ is slightly smaller than the force-free radial field with $m=1/\Lambda_0$ except at $r=0,a$~Mm where they match.  This agrees with the MHS loops being slightly less bulged than the force-free loops.

In the force-free solution class, $m$ depends on the twist parameter $\alpha$ and the radial flux distribution associated with $k$ (Figure~\ref{compare}), whereas in the non-force-free solution class $\Lambda$ is determined entirely by the temperature of the injected plasma. Since these two parameters are independent, any of the force-free solutions plotted in Figure~\ref{coshloops} could evolve to any of the non-force-free solutions, provided that certain invariants of the ideal MHD equations are conserved and certain boundary conditions are met.  In the corona, conditions nearly correspond to those of ideal MHD, meaning that  evolution from one static state to another must preserve the total flux through the tube, as well as the helicity trapped in the magnetic field to a very good approximation.  The latter condition is met by imposing the same constant value of $\alpha$ on beginning and end states.  The former is guaranteed by the fact that we have normalized each solution so that the total flux is equal in all solutions in Figure~\ref{coshloops}.

Not only must the total magnetic flux and helicity in the loop be conserved, but the boundary distribution of vertical flux $B_z|_{z=\pm L}$ and the distribution of twist $B_{\theta}|_{z=\pm L}$ must be approximately conserved also.  These conditions are physically imposed on the coronal field by the inertia of the heavy plasma in layers beneath, which forbids the axial flux and twist from being significantly redistributed.  As we have discussed, Figure~\ref{compare} (bottom pictures) shows that these distributions are approximately equal in the vicinity of the loops for all examples, force-free and MHS, shown in Figure~\ref{coshloops}, and that there is no difference at all between the distributions for the force-free examples shown in the left pictures of Figure~\ref{coshloops}, nor are there any differences between the these distributions for the MHS examples of the right pictures.  In the domain $\psi\le 1$, where both classes of model are valid, the force-free and MHS distributions of $B_z|_{z=\pm L}$ and $B_{\theta}|_{z=\pm L}$ are approximately the same for the modest values of $k\approx\alpha$ expected in the corona.

Therefore, under the restrictions described so far, each solution in Figure~\ref{coshloops} could in principle be ideally evolved into any other.  
On the other hand, an important ingredient missing from these models is a suitable external magnetic field in the MHS models.  Those presented here have unphysical plasma pressures that distort the external field - an effect that grows with distance from the loop.  Consideration of our MHS magnetic fields outside the boundary flux contour $\psi =1$ is not strictly meaningful because for $\psi >1$, in view of the linear form of Equation~(\ref{plinear}), the plasma pressure $p$ becomes negative, a deficiency shared with the few existing MHS loop flux tube models (Beli\"en et al. 1997a, b, Bellan~2003) whose expressions for the plasma pressure take a linear form.  Likewise,  the 
force-free
external fields
are not useful as models of coronal phenomena outside the loop.  In Figure~\ref{compare}, the unphysical $\alpha=1/4$ curves for $0<r<20$~Mm give a good idea of 
what our $\alpha=1/16$ curves
look like in $0<r<80$~Mm.  In MHS and force-free cases we cannot expect the distant coronal field to wind around a single central axis.  A model combining a loop in MHS or twisted force-free equilibrium with a reasonable force-free external field will necessitate a numerical treatment to cope with the nonlinearities involved, along the lines of the prominence models of Petrie et al.~(2007).  In the meantime, we can interpret the analytical solutions presented here, taking into account the likely influence of the external coronal field.

At the boundary $\psi =1$ the loop is constrained by the external field, near-potential fields or other twisted loops not included
in the models but whose effects on the loop can still be
anticipated.  In reality, while the field inside a loop is relaxing to equilibrium with its newly injected plasma, the force-free field around the loop will react to the changing shape of the loop, moving to fill space vacated by the loop if it narrows, or else being pushed outward by the loop if it widens.  Although the external fields in these linear solutions are not appropriate to the physics of a real external coronal field and plasma, the essential physics can be inferred from the total pressure imposed on the loop by the external field at the boundary.  In particular, at the apex $z=0$ the total pressure on our MHS loop model takes a simple form dependent only on $\Lambda$:

\begin{equation}
|{\bf B}|^2_{\psi =1, z=0} = B_0^2\frac{4+\alpha^2 a^2\cosh (L/\Lambda )}{a^4\cosh^2 (L/\Lambda )} .
\end{equation}

\noindent
This quantity is graphed as a function of $\Lambda$ with $\alpha =1/16$~Mm, $a=4$~Mm and $L=100$~Mm in Figure~\ref{strengths}.  Clearly this pressure increases significantly as the temperature and the scale height $\Lambda$ increases and as the model loop becomes narrower.  The extreme case $\Lambda\rightarrow\infty$,  Bellan's~(2003) model, has a much higher boundary apex pressure, represented by the dotted line, than any of the models in Figures~\ref{coshloops} and \ref{compare}.  On the other hand, Figure~\ref{compare} shows that the equivalent force-free fields with $m=1/\Lambda$ have smaller boundary pressure than their MHS counterparts.  This means if that a force-free solution in a given row of Figure~\ref{coshloops} with a given value of $m$ is injected with plasma with scale height $\Lambda =1/m$, the final MHS equilibrium cannot be as narrow as the one on the same row of Figure~\ref{coshloops} because its boundary pressure is too strong.  If the loop were to become so narrow, there would be a total pressure imbalance at the boundary, and so the boundary would move outward until pressure balance were restored.   In the same way, if plasma of arbitrary scale height $\lambda >1/m$ is injected, the correspondingly narrower equilibrium could not be attained because the total loop boundary pressure would be too large for the external field total pressure.  Thus it is clear that an expanded force-free loop in Figure\ref{coshloops}, on being injected with low-$\beta$ plasma, must attain final equilibrium with higher expansion than either its MHS counterpart in Figure~\ref{coshloops} or the MHS solution corresponding to the scale height $\Lambda$ of the injected plasma, neither of which solutions is of uniform cross section at EUV temperatures.  If the MHS solution corresponding to $\Lambda$ is much narrower than the initial force-free solution then the actual final MHS state must have much smaller total boundary pressure than this solution and must therefore be much more expanded.  Therefore we can only conclude that plasma injection cannot explain the uniform cross-sectional profiles that we see in images of loops.

\section{Discussion}
\label{discussion}

We have revisited the long-standing mystery of why loop-shaped coronal emission signatures have constant cross-sectional width, while models lead us to expect coronal flux tubes to have width increase with height.  This work has concentrated on the effects of a uniform gravitational field on the magnetic field and plasma distribution of a moderately twisted coronal loop.  Bellan~(2003) argued that an initially empty twisted loop, on being injected with plasma in the absence of gravity, must evolve to a thin equilibrium state of constant cross-section.  We examined equilibrium state including effects of gravitational stratification, motivated by the fact that measurements of widths of EUV loops at about 1~MK are consistent with constant cross section even though their hydrostatic stratification scales are comparable to or smaller than the sizes of the loops themselves.  Introducing gravity into the modeling results in equilibria, with MHS forces, of non-constant cross section in general, with near-constant cross section only when the loop is hot enough for the stratification length scale to be significantly larger than the size of the loop.  Taking into account the influence of the external coronal field, the injection of plasma into an empty, expanded loop could not result in the loop becoming significantly narrower, certainly not narrow enough to have nearly constant cross-section.  Therefore, while we have not solved the mystery of the constant-width loop, we have decisively eliminated plasma forces as a possible cause for this phenomenon for large ($\ge $~100,000~km) at typical coronal ($\approx $~1~MK) temperatures.

The MHS solutions applied here are restricted to a special linear class of solutions, which correspond to a specific radial distribution of injected plasma.  Generally plasma injection into one of these linear force-free loops would result in a highly nonlinear MHS state not describable in closed analytical form.  Also we neglected the influence of surrounding fields, following many previous authors in taking the loop boundary as the boundary of the problem itself.  The value of the special linear MHS solutions applied here lies in the clarity with which they bring out the essential physical consequences of the plasma flows into the force-free loops and of the plasma forces on the final equilibrium.

The theoretical basis for expecting coronal flux tubes to expand is clear.  Near-constant flux tube cross sections are inconsistent with solenoidal (${\bf\nabla}\cdot{\bf B}=0$) field
strengths decreasing with height.  Field strength and flux tube width are inversely proportional for any solenoidal field, and so a field whose strength decreases with increasing altitude should have a corresponding increase in flux tube width.  A dipole field strength falls off with increasing radius
as $1/r^3$ while higher-order potential fields including
active-region loops should fall off more steeply still.
Thus even taking moderate macroscopic nonlinear
currents into account, loop widths should generally
increase with height as quickly as $r^3$ at least.

This leads us
to question whether the loop-shaped emission patterns of constant width really represent flux
tubes.  Perhaps these patterns delineate not flux tubes but loci of magnetic field diffusions
that, locally to very narrow regions, heat successive narrow
loop strands and accelerate them from these regions during the
field reconnections (Petrie~2006). Dissipations of spontaneous
current sheets are expected to occur along flux tube boundaries
(Parker~1994) and may well trace out magnetic field trajectories.
If diffusion region sizes, accelerations due to field diffusions, and
plasma cooling times are uncorrelated with height in the corona,
then diffusion processes should produce emission patterns of approximately
constant width.  It is hoped that coronal magnetic field measurements
(Lin et al. 2004; Tomczyk et al. 2004) will attain sufficiently
high spatial resolution for us to check directly whether
field strengths within loops whose emission profiles have constant cross-section decrease significantly with height.

There are many difficulties in determining whether
emission patterns really do represent flux tubes and, if so,
of measuring the typical cross-sectional area variation with length
of these flux tubes. Furthermore a loop is highly unlikely to have
a simple cross-sectional shape in the solar atmosphere or to maintain
a single shape along its entire length (Parker~1994, Gudiksen \&
Nordlund~2005, L\'opez Fuentes et al.~2006) and this may greatly influence its appearance from
certain angles.  The identification and characterization of individual loops will hopefully be aided by NASA's  STEREO mission.


\acknowledgements
I thank the referee for helpful and constructive comments, and in particular for stressing the importance of the effects of external fields.  I also thank Paul Bellan and Eve Stenson for illuminating discussions.  Some of this work was conducted while the author was a participant in the National Aeronautics and Space Administration (NASA) Postdoctoral Program at Goddard Space Flight Center, and was based at National Solar Observatory, Tucson.

\end{document}